\documentclass[aps,twocolumn,preprintnumbers,amsmath,amssymb,nofootinbib,superscriptaddress,notitlepage]{revtex4-1}

\usepackage{epsfig}
\usepackage[utf8]{inputenc}

\begin{document}

\title{Three-Body $J^P = 0^+,1^+,2^+$ $B^* B^* \bar{K}$ Bound States}

	\author{M. Pavon Valderrama}\email{mpavon@buaa.edu.cn}
\affiliation{School of Physics and Nuclear Energy Engineering, \\
International Research Center for Nuclei and Particles in the Cosmos and \\
Beijing Key Laboratory of Advanced Nuclear Materials and Physics, \\
Beihang University, Beijing 100191, China} 

\date{\today}


\begin{abstract} 
\rule{0ex}{3ex}
Three body systems with short-range interactions display universal features
that have been extensively explored in atomic physics,
but apply to hadron physics as well.
Systems composed of two non-interacting identical particles
(species H) of mass $M$ and a third particle (species P) of mass $m$
that interacts attractively with the other two
have the property that they are more likely to bind
for larger values of the mass ratio $M/m$.
This is particularly striking if the HHP system is in P-wave
(while the interacting pair is in S-wave),
in which case one would not normally expect the formation
of a three body state.
If we assume that the $B^* \bar{K}$ binds to form the $B^*_{s1}$ heavy meson
and notice that the mass ratio of the $B^*$ to $\bar{K}$ is $M /m = 10.8$,
concrete calculations indicate that
there should be a three body $B^* B^* \bar{K}$
bound state between $30-40\,{\rm MeV}$ below the $B^*_{s1} B^*$ threshold.
For the $\Xi_{bb} \Xi_{bb} \bar{K}$ system the mass imbalance is about
$M /m = 20.5$ and two bound states are expected to appear,
a fundamental and an excited one located at $50-90$ and $5-15\,{\rm MeV}$
below the $\Xi_{bb} \Omega^*_{bb\frac{1}{2}}$ threshold
(where $\Omega^*_{bb\frac{1}{2}}$ denotes
the $\Xi_{bb} \bar{K}$ bound state).
We indicate the possibility of analogous P-wave three body bound states
composed of two heavy baryons and a kaon or antikaon and investigate
the conditions under which the Efimov effect could appear
in these systems.

\end{abstract}

\maketitle

\section{Introduction}

The three boson system in the unitary limit shows a geometric spectrum
of shallow bound states, the Efimov effect~\cite{Efimov:1970zz}.
In this limit there is a geometric tower of three body bound states
for which the ratio of the binding energies of the $n$-th
and $(n+1)$-th excited state is given by $E_{n} / E_{n+1} \simeq 512$,
a prediction that has been experimentally confirmed
with cesium atoms~\cite{Kraemer:2006}.
The existence of a geometric spectrum extends
to other three body systems where only two of the three particles
interact resonantly~\cite{Efimov:1970zz,Braaten:2004rn,Naidon:2016dpf}.
Recently it has been found that a similar geometric spectrum
might also arise in specific two-body hadronic systems,
for instance $\Sigma_c \bar{D}^* - \Lambda_{c1} \bar{D}$ and
$\Sigma_c \Xi_b' - \Lambda_{c1} \Xi_b$~\cite{Geng:2017hxc}.
Particularly interesting are three body systems with a mass imbalance
in which we have two-identical particles of the species H with mass $M$
and a third particle of the species P with mass $m$.
When the HH subsystem is non-interacting and the HP subsystem is resonant,
the three body system will eventually display a geometrical Efimov-like
spectrum if the ratio $M / m$ is
big enough~\cite{Braaten:2004rn,Naidon:2016dpf},
as observed in experiments with lithium and cesium atoms~\cite{Pires:2014zza}.
This is not such a surprise if the system is in S-wave,
where there will always be a geometric spectrum~\footnote{
In the absence of spin, isospin or other quantum numbers
that might translate into numerical factors diminishing attraction.}.
But the cases in which the system is in P-wave or higher is
much more interesting, as they are less trivial.
For P-wave this happens for $M / m \geq 13.6$
while for D-wave the threshold is $M / m \geq 38.6$~\cite{Helfrich:2011ut}.
Kartavtsev and Malykh also made
the remarkable discovery~\cite{Kartavtsev:2006aa}
that in the P-wave case there is a universal three body state
for $M / m \geq 8.176$ and a second one for $M / m \geq 12.917$.
By universal it is meant that the binding energies of these three-body
bound states depend only on the two-body binding energy.

The bottom-line is that three-body systems with large mass imbalances
are more likely to bind.
This is particularly interesting in view of the recent renaissance of
heavy hadron spectroscopy triggered by the discovery of
the $X(3872)$~\cite{Choi:2003ue} (which has been theorized
to be a shallow two-body bound state~\cite{Tornqvist:2003na,Voloshin:2003nt,Braaten:2003he}).
The $D K$ and $D^* K$ systems display a strong s-wave attraction
that generates a bound state at about $45\,{\rm MeV}$
below threshold~\cite{Kolomeitsev:2003ac,Guo:2006fu,Guo:2006rp,Guo:2009ct,Altenbuchinger:2013vwa}.
These bounds states are suspected to be the $D^*_{s0}(2317)$ and $D^*_{s1}(2460)$
charmed mesons, partly because the $D K$ and $D^* K$
bind at the right location partly because of other reasons,
like the fact that the masses of the $D^*_{s0}$/$D^*_{s1}$ are
similar to (instead of markedly heavier than)
those of the $D_0$/$D_1$ charmed mesons or the analysis 
of the $D^*_{s0}$/$D^*_{s1}$ wave function
from lattice data~\cite{Torres:2014vna,Bali:2017pdv}.
Owing to heavy flavor symmetry this idea extends to the $B {\bar K}$
and $B^* {\bar K}$ cases, which form the $B^*_{s0}$ and $B^*_{s1}$ mesons.
Last, if we consider heavy antiquark-diquark symmetry
(HADS)~\cite{Savage:1990di,Hu:2005gf,Guo:2013xga}
then a new set of
bound states involving $\Xi_{cc} \bar{K}$, $\Xi_{cc}^* \bar{K}$,
$\Xi_{bb} \bar{K}$ and $\Xi_{bb}^* \bar{K}$ should appear,
which we will call the $\Omega^*_{cc\frac{1}{2}}$, $\Omega^*_{cc\frac{3}{2}}$,
$\Omega^*_{bb\frac{1}{2}}$ and $\Omega_{bb\frac{3}{2}}^*$
in analogy with the $D^*_{s0}$, $D^*_{s1}$ notation.
Owing to the slightly larger reduced masses the binding energies are also a
bit bigger than in the $DK$ and $D^*K$ cases, of the order of
$60-70\,{\rm MeV}$~\cite{Guo:2011dd}.
In particular if we consider the $B^*B^* \bar{K}$ and
$\Xi_{bb} \Xi_{bb} \bar{K}$ / $\Xi_{bb}^* \Xi_{bb}^* \bar{K}$
the masses imbalances are remarkable, $10.8$ and $20.5$ respectively
(where for the mass of the doubly bottom baryons we have used
the lattice QCD determination of Ref.~\cite{Lewis:2008fu}).
This points out to the possibility of P-wave three body bound states.
Concrete calculations show that this is indeed the case for the bottom hadrons,
with $B^*B^* \bar{K}$ binding about $30-40\,{\rm MeV}$
below the $B_{s1}^* B^*$ threshold, where it is interesting to notice that
this state can also be predicted in a two-body description involving
the $B_{s1}^* B^*$ mesons interacting by means of a
one antikaon exchange potential~\cite{SanchezSanchez:2017xtl}.
For the $\Xi_{bb} \Xi_{bb} \bar{K}$ / $\Xi_{bb}^* \Xi_{bb}^* \bar{K}$
two bound states appear,
a shallow one with a binding of $5-15\,{\rm MeV}$ below the
$\Omega_{bb\frac{1}{2}}^* \Xi_{bb}$ / $\Omega_{bb\frac{3}{2}}^* \Xi_{bb}^*$
threshold and a second one at $50-90\,{\rm MeV}$.
Meanwhile the charmed mesons and doubly charmed baryons
are unlikely to bind in P-wave as a consequence of
the insufficient mass imbalance.
Yet they will likely bind in S-wave~\cite{SanchezSanchez:2017xtl},
as happen in other S-wave, mass-imbalanced three hadron systems
like the $\rho D^*\bar{D}^*$~\cite{Bayar:2015oea},
$\rho B^*\bar{B}^*$~\cite{Bayar:2015zba} and
$K D^*\bar{D}^*$~\cite{Ren:2018pcd}
systems. 

This idea also applies to other P-wave HHP hadron systems
with large masses imbalances.
If we consider the H hadron to be a bottom baryon and the P hadron
to be a kaon or antikaon, the HP interaction is
of a Weinberg-Tomozawa type and in a few cases
might be strong enough as to bind the HP subsystem~\cite{Lu:2014ina}.
If this is the case, this will likely imply the existence of HHP bound states.
At this point the natural question arises of whether the P-wave Efimov effect
will be present in these systems if the HP interaction is resonant.
The answer is negative for two bottom baryon plus a kaon/antikaon
because the mass imbalance is not large enough.
However from HADS~\cite{Savage:1990di}
we expect the existence of doubly heavy tetraquark partners of
the heavy baryons.
If these doubly heavy tetraquarks are stable
they will be the perfect candidates.
In this regard we notice that the recent discovery of a doubly charmed baryon
by the LHCb~\cite{Aaij:2017ueg} strongly points towards the stability of
doubly heavy tetraquarks
in the bottom sector~\cite{Karliner:2017qjm,Mehen:2017nrh}.

The manuscript is structured as follows: after the introduction,
we explain the Faddeev equations for the $B^* B^* \bar{K}$ system
in Section \ref{sec:faddeev}.
We discuss the conditions for the appearance of the P-wave Efimov effect
in Section \ref{sec:efimov}.
Then we show the predictions for $B^* B^* \bar{K}$ and
$\Xi_{bb} \Xi_{bb} \bar{K}$ / $\Xi_{bb}^* \Xi_{bb}^* \bar{K}$ P-wave three body
states in Section \ref{sec:predictions}.
Finally we present our conclusions at the end. 

\section{Faddeev Equations for the HHP System in P-wave}
\label{sec:faddeev}

Here we present the Faddeev equations for solving the HHP bound state
problem for the P-wave case.
This is done for the particular case of contact interactions.
If the HP system is $B^* \bar{K}$, $\Xi_{bb} \bar{K}$ or $\Xi_{bb}^* \bar{K}$,
the binding momentum of the antikaon lies on the vicinity of $200\,{\rm MeV}$.
This is comparable with the mass of the antikaon, $m_K = 495\,{\rm MeV}$,
which means that relativistic kinematics might have a moderate impact
on the calculations.
For this reason we will present first the standard non-relativistic Faddeev
equations and then we will explain how to include corrections coming
from the relativistic antikaon kinematics.
Concrete calculations show that though relativistic corrections 
are not negligible, they are not required at the level of accuracy
at which the HHP bound states can be computed now.

\subsection{The Equations}

We begin with the Faddeev decomposition of the HHP wave functions
\begin{eqnarray}
  \Psi_{3B} &=&
  \left[\phi(\vec{k}_{23}, \vec{p}_1) - \phi(\vec{k}_{31}, \vec{p}_2)\right]\,
  | 1 \otimes \frac{1}{2} \rangle_{1/2} \, , \label{eq:Psi_3B}
\end{eqnarray}
with particles $1$, $2$ and $3$ corresponding to species H, H and P
(particles $1$ and $2$ are identical).
This decomposition indicates that the HH subsystem is antisymmetric
in the spatial coordinates, which implies it has odd orbital
angular momentum $L_{12} = 1, 3, 5, \dots$ with the $L_{12} = 1$
component dominant at low momenta.
It also assumes that there is no interaction in the HH subsystem,
a hypothesis that we will review in a few lines.
The Jacobi momenta $\vec{k}_{ij}$ and $\vec{p}_k$ are defined as usual:
\begin{eqnarray}
  \vec{k}_{ij} &=& \frac{m_j \vec{k}_i - m_i \vec{k}_j}{m_i + m_j} \, , \\
  \vec{p}_{k} &=& \frac{1}{M_T}\,\left[ (m_i + m_j)\,\vec{k}_k -
    m_k\,(\vec{k}_i + \vec{k}_j) \right] \, , 
\end{eqnarray}
with $m_1$, $m_2$, $m_3$ the masses of particles $1$, $2$, $3$ (we take
$m_1 = m_2 = M$ and $m_3 = m$), $M_T = m_1 + m_2 + m_3$ the total mass
and $ijk$ an even permutation of $123$.
The ket refers to the isospin wave function of the system in the notation
\begin{eqnarray}
  | I_{12} \otimes I_3 \rangle_{I_T} \, ,
\end{eqnarray}
where $I_{12}$ is the isospin of particles $1$ and $2$, $I_3$ the isospin
of particle $3$ and $I_T$ the total isospin.
The choice $I_{12} = 1$, $I_T = \frac{1}{2}$ is the combination
with the biggest overlap into the $I=0$ channel of the HP subsystem,
where the $B_{s1}^*$ bound state is expected to happen.
The spin wave function is not explicitly indicated: the $B^*$'s are bosons,
their isospin wave function is symmetric and the spatial wave function
is antisymmetric, from which we deduce that $S_{12} = 1$.
The coupling of the spin of the $B^*$ mesons with their orbital angular
momentum $L_{12} = 1$ leads to the conclusion that the quantum numbers
of the three-body bound states are $J^{P} = 0^{+}$, $1^{+}$ and $2^{+}$.
The same logic applies if we consider the $\Xi_{bb}$ and $\Xi_{bb}^*$ baryons,
though in this case we have spin $\frac{1}{2}$ and $\frac{3}{2}$ fermions:
we end up with $S_{12} = 1$ or $S_{12} = 1, 3$ for the spin wave function,
where the quantum numbers of the states are $J^{P} = 0^{+}$, $1^{+}$ and $2^{+}$
for $\Xi_{bb} \Xi_{bb} \bar{K}$ and $J^{P} = 0^{+}$, $1^{+}$, $2^{+}$, $3^{+}$
and $4^{+}$ for $\Xi_{bb}^* \Xi_{bb}^* \bar{K}$.

The interaction in the HH subsystem is not zero
but it is expected to be small.
The orbital angular momentum is $L_{12} \geq 1$,
which effectively suppresses the short-range components of
the interaction (rho- and omega-exchange, for instance).
If we consider the long-range interaction instead, which is given by
the one pion exchange (OPE) potential, we notice that
the isospin of the HH subsystem is $I_{12} = 1$.
The OPE potential is known to be weak in the isovector configurations of
the $B^* B^*$ and $B^* \bar{B}^*$ systems, as already pointed out
in the seminal work of T\"ornqvist~\cite{Tornqvist:1993ng}.
For the $\Xi_{bb} \Xi_{bb}$ system the strength of OPE is $1/9$ of that of
the $B^* B^*$ system, a result which can derived from HADS
(see Ref.~\cite{Liu:2018bkx} for instance). 
That is, OPE is suppressed for the HH configurations we are considering here.
This in turn implies that the HH interaction is likely to be a perturbative
effect that we can neglect owing to the exploratory character of
the current calculations.

The HP interaction is of a short-range type. We can write it as
\begin{eqnarray}
  V_{23} = C\,g(k)\,g(k') \, , \label{eq:pot-nr}
\end{eqnarray}
where $k$, $k'$ are the initial and final relative momenta of
particles $2$ and $3$, while $g(k)$ is the regulator function we are using.
From this potential the T-matrix is given by the ansatz
\begin{eqnarray}
  T_{23}(Z) &=& \tau_{23}(Z) \, g(k) g(k')  \, ,
\end{eqnarray}
where $Z$ refers to the energy.
The coupling $C$ is determined from the condition that $\tau_{23}(Z)$
has a pole at the location of the $B_{s1}^*$ strange-bottom meson. 
For the Faddeev component of the wave function there is the well-known ansatz
\begin{eqnarray}
  \phi(\vec{k}, \vec{p}) &=&
  \frac{g(k)}{Z - \frac{k^2}{2 \mu_{23}} - \frac{p^2}{2 \mu_1}}\,
  a_1(p) Y_{1 m}(\hat{p}) \, , 
\end{eqnarray}
where $Y_{1 m}$ is a spherical harmonic and $\mu_{ij}$ and $\mu_k$
are reduced masses defined as
\begin{eqnarray}
  \frac{1}{\mu_{ij}} &=& \frac{1}{m_i} + \frac{1}{m_j} \, , \\
  \frac{1}{\mu_{k}} &=& \frac{1}{m_k} + \frac{1}{m_i + m_j} \, .
\end{eqnarray}
The wave function is fully determined by $a_1(p)$,
for which the Faddeev equations can be reduced to~\footnote{
Notice that the integral equation for $a_1(p_1)$ does not include
the integration on the $\hat{p}_1$ angular variable.
The reason is that it is not required: the integrand on the right hand side
of Eq.~(\ref{eq:faddeev-a1}) depends only on the angle
between $\vec{p}_1$ and $\vec{p}_2$, which is already
taken care of by the integration on the $\hat{p}_2$ angular variable.
}
\begin{eqnarray}
  a_1(p_1) &=& -\frac{3}{4}\,
  \tau_{23}(Z_{23}) \, \int \frac{d^3 \vec{p}_2}{(2 \pi)^3}
  B^{1}_{12}(\vec{p}_1, \vec{p}_2)\, a_1({p}_2) \, , \nonumber \\
  \label{eq:faddeev-a1}
\end{eqnarray}
where $Z_{23} = Z - \frac{p_1^2}{2 m_1}\,\frac{M_T}{m_2 + m_3}$ and
$B^1_{12}$ is given by
\begin{eqnarray}
  B^1_{12}(\vec{p}_1, \vec{p}_2) &=&
  \frac{g(q_1) g(q_2)}
       {Z - \frac{p_1^2}{2 m_1} - \frac{p_2^2}{2 m_2} -
         \frac{p_3^2}{2 m_3}} \, P_1(\hat{p}_1 \cdot \hat{p}_2) \, ,
\end{eqnarray}
where $P_1(x)$ is a Legendre polynomial.
We have that $\vec{p}_1 + \vec{p}_2 + \vec{p}_3 = 0$ and
\begin{eqnarray}
  \vec{q}_i = \frac{m_j \vec{p}_k - m_k \vec{p}_j}{m_j + m_k} \, ,
\end{eqnarray}
with $ijk$ an even permutation of $123$.
Once we have all the pieces we can solve the eigenvalue equation
by discretization and obtain the energy of the bound states.

\subsection{Inclusion of Relativistic Effects}

Previously we have considered the kaons to behave non-relativistically.
The binding momentum of the typical HP bound states
is close to $200\,{\rm MeV}$.
This indicates that relativistic corrections to the kaon kinematics
might have a moderate impact on the three body binding.
The derivation of relativistic Faddeev equations for systems
with contact-range interactions is not unique,
a situation which is analogous to what happens
in the two-body system~\cite{Kadyshevsky:1967rs,Gross:1969rv,Yaes:1971vw,Woloshyn:1974wm,Ramalho:2001pd}.
Here we choose to follow the prescription of Garcilazo and
Mathelisch~\cite{Garcilazo:1984rx,Mathelitsch:1986ez},
which reproduces the Kadyshevsky equation~\cite{Kadyshevsky:1967rs}
for the two-body sector.
We adapt this prescription to the problem at hand,
where the only non-relativistic
particle is the kaon and the mass of the heavy hadrons
is considerably larger than the kaon energy.
This amounts to the following change in the two-body propagator
for the calculation of the two-body T-matrix
\begin{eqnarray}
  \frac{1}{Z - \frac{p_2^2}{2 m_2} - \frac{p_3^2}{2 m_3}} \to
  \frac{m_3}{\omega_3(p_3)}\,
  \frac{1}{Z - \frac{p_2^2}{2 m_2} - \epsilon_3(p_3)} \, ,
  \label{eq:G2-rel}
\end{eqnarray}
plus the analogous modification for $B_{12}^1$
\begin{eqnarray}
  B^{1}_{12}(\vec{p}_1, \vec{p_2}) \to
  \frac{m_3}{\omega_3(p_3)} \,
  \frac{g(q_1) g(q_2)\,P_1(\hat{p}_1 \cdot \hat{p}_2)}
       {Z - \frac{p_1^2}{2 m_1} - \frac{p_2^2}{2 m_2} -
         \epsilon_3(p_3)} \,  \, ,
\end{eqnarray}
with $\omega_3(q) = \sqrt{m_3^2 + q^2}$, where $m_3 = m_K$
is the kaon energy and $\epsilon_3(q) = \omega_3(q) - m_3$.
The advantage of this prescription is that the changes are easy to implement
from the computational point of view.

Besides kinematics, another relativistic effect is that
the HP interaction is of a Weinberg-Tomozawa type,
which is not momentum independent as previously assumed.
The correct momentum dependence is indeed
\begin{eqnarray}
  V_{23} = C\,\left[ \frac{\omega_3(k) + \omega_3(k')}{2 m_3} \right]\,
  g(k)\,g(k') \, . \label{eq:pot-r}
\end{eqnarray}
This potential can also be rewritten as
\begin{eqnarray}
  V_{23} = C\,\left[ 1 + f(k) + f(k') \right]\,g(k)\,g(k')\, \, ,
\end{eqnarray}
where $f(k) = (\omega_3(k) - m_3) / 2 m_3$.
The T-matrix for this potential admits a well-known ansatz
\begin{eqnarray}
  T_{23}(Z) &=& g(k) g(k')\,
  \Big[ \tau_{23}^{A}(Z) + \tau_{23}^B(Z) (f(k) + f(k'))
    \nonumber \\ && \qquad \quad
    + \, \tau_{23}^C(Z) f(k) f(k')
    \Big] \, ,   
\end{eqnarray}
plus the following ansatz for the Faddeev component
\begin{eqnarray}
  \phi(\vec{k}, \vec{p}) =
  \frac{g(k)\,\left[ a_1(p) + b_1(p)\,f(k)
    \right]}{Z - \frac{p_1^2}{2 m_1} - \frac{p_2^2}{2 m_2} - \epsilon(p_3)}\,
  \, Y_{1m}(\hat{p}) \, , \nonumber \\
\end{eqnarray}
where $p_1 = p$, $p_2 = k - m_2 \, p / (m_2 + m_3)$ and
$p_3 = -k - m_3 \, p / (m_2 + m_3) $.
This leads to a different set of Faddeev equations:
\begin{eqnarray}
  a_1(p_1) &=&  -\frac{3}{4}\,\int \frac{d^3 \vec{p}_2}{(2 \pi)^3}
  \left[ \tau_{23}^A + \tau_{23}^B\,f(q_1) \right]
  B^{1}_{12}(\vec{p}_1, \vec{p}_2)\, \nonumber \\ && \qquad
  (a_1({p}_2) + b_1(p_2) \, f(q_2)) \, , \\
  b_1(p_1) &=&  -\frac{3}{4}\,\int \frac{d^3 \vec{p}_2}{(2 \pi)^3}
  \left[ \tau_{23}^B + \tau_{23}^C \,f(q_1) \right]
  B^{1}_{12}(\vec{p}_1, \vec{p}_2)\,
  \nonumber \\ && \qquad
  (a_1({p}_2) + b_1(p_2)\, f(q_2)) \, ,
\end{eqnarray}
where the $\tau_{23}^{(A,B,C)}$ components of
the T-matrix are evaluated at $Z = Z_{23}$.

\section{The Efimov Effect in the HHP System}
\label{sec:efimov}

Now we consider the Faddeev equations in the unitary limit, i.e.
when the binding energy of the HP state approaches zero.
For $Z \to 0$ and momenta $p_1$, $p_2$ well below the cut-off
we have the simplifications
\begin{eqnarray}
  \tau_{23}(Z_{23}) &\to& -
  \frac{2 \pi}{\mu_{23}}\,\sqrt{\frac{\mu_{23}}{\mu_1}}\frac{1}{p_1} \, , \\
  \int \frac{d^2 \hat{p_2}}{4 \pi} B_{12}^1 &\to&
  + \frac{m}{p_1 p_2} Q_1(\frac{M+m}{2 M} \frac{p_1^2 + p_2^2}{p_1 p_2}) \, ,
\end{eqnarray}
where $Q_1(z)$ is a Legendre function of the second kind
\begin{eqnarray}
  Q_1(z) = \frac{z}{2}\,\log{\frac{z+1}{z-1}} - 1 \, .
\end{eqnarray}
If we ignore the purely polynomial terms in $p_1$ and $p_2$,
we end up with the equation
\begin{eqnarray}
  p_1^3 a(p_1) &=& \frac{3}{4}\,\frac{1}{\pi}\,\sqrt{\frac{\mu_1}{\mu_{23}}}\,
  {\left(\frac{M+m}{2 M}\right)}^2
  \int_0^{\infty} dp_2\,a(p_2)\, \nonumber \\
  &\times& 
  (p_1^2 + p_2^2)\,\log{\left( \frac{p_1^2+p_2^2 + \frac{2M}{M+m} p_1 p_2}
  {p_1^2+p_2^2 - \frac{2M}{M+m} p_1 p_2 } \right)} \, . \nonumber \\
\end{eqnarray}
If we assume a solution of the type $a(p) = b(p)/p^3$ with $b(p) = p^s$,
we find the eigenvalue equation
\begin{eqnarray}
  1 &=& \frac{3}{4}\,\frac{1}{\pi}\,\sqrt{\frac{\mu_1}{\mu_{23}}}\,
  {\left(\frac{M+m}{2 M}\right)}^2 \nonumber \\
  &\times& \int_0^{\infty} dx\, x^{s-3} (1+x^2) \,
  \log{\left( \frac{1+x^2 + \frac{2M}{M+m} x}
    {1+x^2 - \frac{2M}{M+m} x } \right)} \nonumber \\
  &=& \frac{3}{4}\,I_E^1(s)\, .
\end{eqnarray}
The integral $I_E^1(s)$ is analytically solvable~\cite{Helfrich:2011ut}
\begin{eqnarray}
  I_E^1(s) = \,\frac{1}{2 \sin^2{\alpha} \cos{\alpha}}\,
  &\Big[& 
    \frac{1}{i s - 1}\,\frac{\sin{\left[ (i s - 1) \alpha \right]}}
         {\cos{\left[ (i s - 1) \frac{\pi}{2} \right]}} \nonumber \\
         &+&
         \frac{1}{i s + 1}\,\frac{\sin{\left[ (i s + 1) \alpha \right]}}
              {\cos{\left[ (i s + 1) \frac{\pi}{2} \right]}} \Big] \, ,
  \nonumber \\
\end{eqnarray}
where $\alpha$ is
\begin{eqnarray}
  \alpha = {\rm asin}{(\frac{1}{1+\delta})} \, ,
\end{eqnarray}
with $\delta = m/M$ the inverse of the mass imbalance.
For $M/m \geq 20.587$ the eigenvalue equation admits complex solutions of
the type $s = \pm i s_1$, indicating the existence of an Efimov geometric
spectrum.

\begin{table}[ttt]
  \begin{tabular}{|cccccccc|}
    \hline
    HHP & $C_{WT}$ & $B_{HP}$($\rm MeV$) & $I_T$ & $I_{12}$ & $c_I$ & $(M/m)$ & $(M/m)_{\rm crit}$  \\
    \hline \hline
    $NN\bar{K}$ & $-3$ & $8$ & $\frac{1}{2}$ & $1$ &
    $\frac{3}{4}$ & $1.9$ & $20.6$ \\
    $B^* B^* K$ & $-2$ & $60-70$ & $\frac{1}{2}$ & $1$ &
    $\frac{3}{4}$ & $10.8$ & $20.6$ \\
    $\Xi_{bb} \Xi_{bb} \bar{K}$ & $-2$ & $60-70$ & $\frac{1}{2}$ & $1$ &
    $\frac{3}{4}$ & $20.5$ & $20.6$ \\
    $\Xi_b' \Xi_b' K$ & $-2$ & N/A & $\frac{1}{2}$ & $1$ &
    $\frac{3}{4}$ & $12.0$ & $20.6$ \\
    $\Sigma_b \Sigma_b \bar{K}$ & $-3$ & N/A & $\frac{1}{2}$ & $1$ &
    $\frac{2}{3}$ & $11.7$ & $24.5$ \\
    &  & & $\frac{3}{2}$ & $2$ & $\frac{5}{6}$ & $11.7$ & $17.7$ \\
    $\Omega_b \Omega_b K$ & $-2$ & N/A & $\frac{1}{2}$ & $0$ &
    $1$ & $12.1$ &$13.6$ \\
    \hline
  \end{tabular}
  \caption{P-wave HHP three body systems with large mass
    imbalances where the Weinberg-Tomozawa interaction of
    the HP subsystem might be able to produce binding.
    The relative strength of the Weinberg-Tomozawa is denoted by $C_{WT}$.
    This leads to the coupling $C = C_{WT} / 2 f_{\pi}^2$ in the potentials
    of Eqs.~(\ref{eq:pot-nr}) and (\ref{eq:pot-r}).
    The approximate binding energy --- if known --- of
    the HP system is shown in the column $B_{HP}$:
    the $N\bar{K}$ value is taken from Ref.~\cite{Hyodo:2007jq},
    the $B^* \bar{K}$ from Ref.~\cite{Guo:2011dd} and
    the $\Xi_{bb} \bar{K}$ value is deduced from HADS.
    For the masses of the experimentally observed heavy hadrons
    we use the isospin average of the values listed
    in the PDG~\cite{Patrignani:2016xqp}, i.e.
    $M(B^*) = 5325\,{\rm MeV}$, $M(\Xi_b') = 5935\,{\rm MeV}$,
    $M(\Sigma_b) = 5813\,{\rm MeV}$ and $M(\Omega_b) = 6046\,{\rm MeV}$.
    For the mass of the doubly bottom baryon $\Xi_{bb}$ 
    we use the central value of the lattice calculation
    of Ref.~\cite{Lewis:2008fu}, i.e.
    $M(\Xi_{bb}) = 10127\,{\rm MeV}$. 
    We consider the HHP and HP systems to have isospin $I_T$ and $I_{12}$,
    resulting in the isospin factor $c_I$.
    If the HP system happens to bind near the threshold and
    the mass imbalance $(M/m)$ of the $H$ hadron and
    the $P$ pseudo Nambu-Goldstone
    boson is larger than the critical value $(M/m)_{\rm crit}$,
    the HHP system might display the P-wave Efimov effect.
    Even though for the heavy hadrons listed above the mass imbalance
    does not reach the critical value, it is probable
    for these systems to have three body bound states
    as the ones we have computed for the $B^* B^* K$ and
    $\Xi_{bb} \Xi_{bb} \bar{K}$ / $\Xi_{bb}^* \Xi_{bb}^* \bar{K}$ systems.
    From HADS we expect the $\Xi_Q'$, $\Sigma_Q$ and $\Omega_Q$ heavy baryons
    to have doubly heavy tetraquark partners $T_{QQ}$, leading to
    mass imbalances twice as big as the ones listed in this table.
  }
  \label{tab:WT-Efimov}
\end{table}

If we consider the $B \bar{K}$ and $\Xi_{bb} \bar{K}$ cases, 
the existence of a geometrical spectrum is a theoretical possibility
rather than a practical one: these systems are too tightly
bound to show this type of universality.
Yet there are many hadrons in with the Weinberg-Tomozawa interaction
with a kaon or antikaon might result in a bound state~\cite{Jido:2003cb,Magas:2005vu,Hyodo:2006yk,Hyodo:2006kg,Hyodo:2007jq}.
The isospin structure can be different, leading to the eigenvalue equation
\begin{eqnarray}
  1 = c_I\,I_E^1(s) \, ,
\end{eqnarray}
where $c_I$ is an isospin factor that depends on the particular case
under consideration.
A few examples with a strongly attractive Weinberg-Tomozawa term 
include the $\Xi_Q' K$, $\Omega_Q K$ and $\Sigma_Q \bar{K}$~\cite{Lu:2014ina}.
The isospin factors and masses imbalances required for the
P-wave Efimov effect are listed in Table \ref{tab:WT-Efimov},
where the relative strength of the Weinberg-Tomozawa term
has also been included.
For the $\Xi_b' \Xi_b' K$ system the isospin factor is identical
to that of $B B K$ and $\Xi_{bb} \Xi_{bb} \bar{K}$, i.e. $c_I = \frac{3}{4}$.
For the $\Omega_Q \Omega_Q K$ system the isospin factor is $c_I = 1$
and the mass imbalance required for a geometrical spectrum is
the standard $13.6$.
This is to be compared with a mass imbalance of $12.1$
for the $\Omega_b K$ case.
For the $\Sigma_Q \Sigma_Q \bar{K}$ system the isospin factors are $\frac{2}{3}$
and $\frac{5}{6}$ for total isospin $I = \frac{1}{2}$ and $I = \frac{3}{2}$
respectively, which require mass imbalances of $24.5$ and $17.7$.
This limit is however not reached for $Q = b$,
in which case the mass imbalance is $11.7$ 
for both the $\Sigma_b$ and $\Sigma_b^*$.
Notice that the previous HP molecules are expected to have a finite width:
the $\Xi_Q' K$ system can decay into $\Sigma_Q \pi$, while
the $\Sigma_Q \bar{K}$ and $\Omega_Q K$ can both decay into $\Xi_Q' \pi$.
Their corresponding HHP bound states will also have a finite width.

From HADS~\cite{Savage:1990di}
we naively expect the existence of doubly heavy tetraquark $T_{QQ}$ partners of
the $T_Q = \Lambda_Q / \Xi_Q$ and $S_Q = \Sigma_Q / \Xi_Q' / \Omega_Q$
heavy baryons.
The Weinberg-Tomozawa interactions for the doubly heavy tetraquarks
will be identical to those of the heavy baryons,
but their mass imbalances will be about twice as high as
the ones listed in Table \ref{tab:WT-Efimov}.
This means that the $T_{QQ} T_{QQ} P$ system is a possible candidate
for the P-wave Efimov effect in hadronic physics.
However the existence of strongly- and electromagnetically-stable tetraquarks
is not guaranteed, as it depends on their locations being below
the relevant open charm/bottom thresholds~\cite{Karliner:2013dqa}.
In this regard it has been pointed out that the actual location of
the the doubly charmed $\Xi_{cc}^{++}$ baryon, recently observed by
the LHCb collaboration~\cite{Aaij:2017ueg}, suggests
the stability of the doubly bottomed
tetraquarks~\cite{Karliner:2017qjm,Mehen:2017nrh}.

\section{Three Body $B^* B^* \bar{K}$ States}
\label{sec:predictions}

\begin{table}[ttt]
  \begin{tabular}{|c|cc|cc|}
    \hline
    HHP &$B_{2}^{\rm NR}$ & $B^{\rm NR}_3$ & $B^{\rm R}_2$ & $B^{\rm R}_3$ \\
    \hline \hline
    $B^* B^* \bar{K}$ & $57-74$ & $32-42$ & $59-81$ & $32-33$  \\
    \hline
    $\Xi_{bb} \Xi_{bb} \bar{K}$ & $60-83$ & $8-14$ & $66-93$ & $2-14$ \\
    & & $50-90$ &  & $52-83$ \\
    \hline
  \end{tabular}
  \caption{Two and three body binding energies in MeV for the $B^* B^* \bar{K}$
    and $\Xi_{bb} \Xi_{bb} \bar{K}$ / $\Xi_{bb}^* \Xi_{bb}^* \bar{K}$
    for different values of the cut-off and depending on the kinematics
    (non-relativistic and relativistic,
    indicated by the superscripts NR and R).
    The two body binding energy $B_2$ refers to the hadron - antikaon system,
    while the three body binding energy $B_3$ is computed with respect
    to the two body binding threshold, i.e. with respect to $(2 M + m - B_2)$
    with $M$ the mass of the hadron and $m$ the mass of the antikaon.
    We make no difference between the $\Xi_{bb} \Xi_{bb} \bar{K}$ /
    $\Xi_{bb}^* \Xi_{bb}^* \bar{K}$ systems as there is no noticeable change
    in the predicted binding energies owing to the similar masses of
    the $\Xi_{bb}$ and $\Xi_{bb}^*$ baryons, $M(\Xi_{bb}) = 10127\,{\rm MeV}$
    and $M(\Xi_{bb}^*) = 10151\,{\rm MeV}$
    according to Ref.~\cite{Lewis:2008fu}.
  }
\label{tab:HHP}
\end{table}

Now we calculate the location of the HHP bound states
for $H = B / \Xi_{bb} / \Xi_{bb}^*$ and $P = \bar{K}$.
For that we need to know the location of the HP bound states,
which is not available experimentally except for the $D K$ and $D^* K$ cases
(unfortunately these two systems do not have a large enough
mass imbalance to form a P-wave bound state).
From heavy flavor symmetry we expect however the $B \bar{K}$ and $B^* \bar{K}$
potential to be identical to that of the $D {K}$ and $D^* {K}$.
The same is true for $\Xi_{bb} \bar{K}$ if we consider HADS.
Besides, the strength of the Weinberg-Tomozawa terms should also
be identical in all these HP systems, thus cementing
the previous conclusions obtained from heavy quark symmetry.
The only difference with the $D {K}$ and $D^* {K}$ systems is
that the reduced mass is a bit larger, approaching the kaon mass
in the $m_Q \to \infty$ limit.
There are a few theoretical calculations of the masses of the aforementioned
HP systems, which are usually inside the $60-70\,{\rm MeV}$
window~\cite{Guo:2006fu,Guo:2006rp,Guo:2011dd}.
Here for consistency we will simply recalculate the location of the HP
partners of the $D {K}$ and $D^* {K}$ systems from the assumption
that the binding energies of the later are known.
We will do two calculations, a non-relativistic and a relativistic one.
For the non-relativistic one we use the potential
\begin{eqnarray}
  V^{\rm NR} = C(\Lambda) g_{\Lambda}(k') g_{\Lambda}(k) \, ,
\end{eqnarray}
where $C(\Lambda)$ is a running coupling constant and
$g_{\Lambda}(k) = e^{-(k^2/\Lambda^2)^{n}}$ is a gaussian regulator with $n = 2$.
For the relativistic calculation
we will include the correct Weinberg-Tomozawa energy dependence
\begin{eqnarray}
  V^{\rm R} = C(\Lambda)\,\frac{\omega_K(k) + \omega_K(k')}{2 m_K}\,
  g_{\Lambda}(k') g_{\Lambda}(k) \, ,
\end{eqnarray}
with $\omega_K(q) = \sqrt{m_K^2 + q^2}$, where we also modify
the two-body propagator in line with Eq.~(\ref{eq:G2-rel}).
We choose the cut-off to float in the $\Lambda = 0.5-1.0\,{\rm GeV}$ window,
i.e. a cut-off around the breakdown scale of the previous description
(which is set by the vector meson mass $m_{\rho} = 0.77\,{\rm GeV}$).
Now if we fix the $D K$ and $D^* K$ binding to $45\,{\rm MeV}$,
in the non-relativistic case we obtain a binding energy of
$57-74\,{\rm MeV}$ and $60-83\,{\rm MeV}$ for the $B^* \bar{K}$
and $\Xi_{bb} \bar{K}$ molecules.
In the relativistic case these numbers increase a bit to $59-81\,{\rm MeV}$
and $66-93\,{\rm MeV}$ respectively.
For the three body system we define the binding energy with respect to
the {\it particle-dimer threshold}, that is, with respect to $2 M + m - B_2$.
This means that the location of the three body bound states is 
\begin{eqnarray}
  M(HHP) = 2 M + m - B_2 - B_3 \, . 
\end{eqnarray}
With this definition the $B^* B^* \bar{K}$ binding energy lies in the range
of $32-42\,{\rm MeV}$ and $32-33\,{\rm MeV}$
for the non-relativistic and relativistic cases, respectively.
For the $\Xi_{bb} \Xi_{bb} \bar{K}$ molecules we find a fundamental and excited
state at $50-90$ and $8-14\,{\rm MeV}$ for non-relativistic antikaons and
$52-83$ and $2-14\,{\rm MeV}$ for relativistic antikaons.
These results are summarized in Table \ref{tab:HHP}.

In the previous calculations we have treated $C(\Lambda)$
as a running coupling constant.
Yet its strength is expected to be given by
\begin{eqnarray}
  C = \frac{C_{WT}}{2 f_{\pi}^2} \, ,
\end{eqnarray}
with $C_{WT} = -2$, where we take the $f_{\pi} = 132\,{\rm MeV}$ normalization.
This suggest a different approach: to treat the coupling $C$ as known
and to choose a cut-off that reproduces the location of
the $D K$ and $D^* K$ poles.
In this case we obtain $\Lambda_{WT} = 0.892 \, {\rm GeV}$
and $0.823\,{\rm GeV}$ for the relativistic and non-relativistic cases.
If we redo the calculations for this {\it privileged} cut-off,
the $B^* \bar{K}$ and $B^* B^* \bar{K}$ lie now at
\begin{eqnarray}
  B_2^{\rm NR} =  71\,{\rm MeV} \quad &\mbox{and}& \quad B_3^{\rm NR} = 40
  \, {\rm MeV} \, ,
  \\
  B_2^{\rm R} =  72\,{\rm MeV} \quad &\mbox{and}& \quad B_3^{\rm R} = 30
  \, {\rm MeV} ,
\end{eqnarray}
depending on whether we are using relativistic or non-relativistic kinematics.
Meanwhile, for the $\Xi_{bb} \bar{K}$ and $\Xi_{bb} \Xi_{bb} \bar{K}$ systems
we have
\begin{eqnarray}
  B_2^{\rm NR} =  78\,{\rm MeV} \quad &\mbox{and}& \quad B_3^{\rm NR} = 9\, / \, 81
  \, {\rm MeV} \, ,
  \\
  B_2^{\rm R} =  79\,{\rm MeV} \quad &\mbox{and}& \quad B_3^{\rm R} = 4\, / \,67 \, {\rm MeV} ,
\end{eqnarray}
where we remind that there is an excited and a fundamental
$\Xi_{bb} \Xi_{bb} \bar{K}$ state.

For comparison purposes we can consider the case of the $\Lambda(1405)$,
which is traditionally considered to be a $N\bar{K}$ bound state.
The strength of the WT term is $C_{WT} = -3$ for this system.
The $\Lambda(1405)$ is known to have a double pole structure~\cite{Jido:2003cb,Magas:2005vu},
which comes from the fact that the $N\bar{K}$ channel mixes
with the $\Sigma \pi$ channel and where the two channels are attractive
enough to generate a pole with the quantum numbers of the $\Lambda(1405)$.
One of the poles is mostly an $N\bar{K}$ bound state. If we ignore
the $\Sigma \pi$ channel, we end up with a standard bound state
which is estimated to be located at $1427\,{\rm MeV}$~\cite{Hyodo:2007jq},
i.e. a binding energy of $8\,{\rm MeV}$.
The cut-offs for which this $\Lambda(1405)$ pole is reproduced
with the formalism presented here are $\Lambda_{WT} = 0.596\,{\rm GeV}$
and $0.571\,{\rm GeV}$ for non-relativistic and relativistic
antikaon kinematics, which are markedly lower than
in the $D K$ and $D^* K$ systems.
The conclusion is that we are not really sure about what is the exact
cut-off to use in the $\Xi_Q' K$, $\Sigma_Q \bar{K}$ and $\Omega_b K$
systems, but we can expect it to be somewhere in between the two values
that we have deduced from the $N \bar{K}$ and $D K$ / $D^* K$ systems.
That is, we expect the cut-off to be somewhere
in the $\Lambda = 0.6-0.9\,{\rm GeV}$ window.
As a matter of fact, for $\Lambda = 0.6\,{\rm MeV}$ all the HP two-body
system of Table \ref{tab:WT-Efimov} ($\Xi_b' K$, $\Sigma_b \bar{K}$
and $\Omega {K}$) bind and the same is true for
the HHP P-wave three-body systems.
For $\Lambda = 0.9\,{\rm GeV}$ the binding energies can in a few cases ---
in particular the $\Sigma_b \bar{K}$ system --- be of the order of a
few hundred ${\rm MeV}$, clearly outside the expected range of
validity of the type of description we are using.
The conclusion is that the spread generated by the cut-off variation
is excessively large: reliable predictions cannot be done
until we find a suitable heavy baryon and kaon/antikaon bound state
from which to fix the contact interaction or the cut-off.
For this reason we will refrain to do concrete predictions
about these systems in this work, except noting their probable existence.

\section{Conclusions}
\label{sec:conclusions}

In this work we have considered the P-wave three body $B^* B^* \bar{K}$ system.
In this system the $B^* \bar{K}$ interaction is strong enough as to generate
a bound state, the $B_{s1}^*$.
In addition the mass imbalance between the $B^*$ and the $\bar{K}$
is remarkable, a feature that points out to the possibility of
P-wave three body bound states.
Concrete calculations indicate that there are indeed P-wave
$B^* B^* \bar{K}$ bound states with quantum numbers $J^P = 0^+$, $1^+$ and $2^+$
located at $30-40\,{\rm MeV}$ below the $B^* B_{s1}$ threshold.
Owing to heavy antiquark-diquark symmetry~\cite{Savage:1990di,Hu:2005gf,Guo:2013xga}
this idea can be easily extended to the $\Xi_{bb} \Xi_{bb} \bar{K}$ system,
where there are two bound states as a consequence of the larger mass imbalance.
In this latter case, the excited and fundamental states are located
about $5-15$ and $50-90\,{\rm MeV}$ below the $\Xi_{bb} \Omega_{bb\frac{1}{2}}^*$
threshold, where $\Omega_{bb\frac{1}{2}}^*$ refers to the theorized
$\Xi_{bb} \bar{K}$ bound state.
In general the antikaon can be treated non-relativistically
in these three body systems, with relativistic corrections
playing a minor role, as we have explicitly checked with calculations.
As a consequence of the isospin and angular momentum of
the $B^* B^*$, $\Xi_{bb} \Xi_{bb}$ and $\Xi_{bb}^* \Xi_{bb}^*$ subsystems,
the possible interaction between the heavy hadrons is expected to
have a very limited impact on the location of the three body states.
It is interesting to notice that the $B^* B^* \bar{K}$ state can
also be predicted in a complimentary two-body description,
in which case we consider a $B^* B_{s1}$ pair interacting by means of
a one antikaon exchange potential~\cite{SanchezSanchez:2017xtl}.
In this interpretation the location of the bound states is a bit more shallow,
about half the binding energy computed here. Nonetheless these figures
are still compatible with the calculations presented here.

This idea could also apply to other HHP systems, particularly if we consider
that the Weinberg-Tomozawa interaction between a hadron and
a pseudo Nambu-Goldstone boson can be strong in some cases.
A few candidate HP systems include the $\Xi_Q' K$, $\Omega_Q K$ and
the $\Sigma_Q \bar{K}$.
If we consider heavy antiquark-diquark symmetry and the observation
that the recent discovery of the $\Xi_{cc}^{++}$
doubly charmed baryon~\cite{Aaij:2017ueg}
probably implies the existence of doubly heavy tetraquarks
in the bottom sector~\cite{Karliner:2017qjm,Mehen:2017nrh},
there is the possibility of a tetraquark-tetraquark-kaon/antikaon
three body system capable of fulfilling the conditions
for the P-wave Efimov effect.

\section*{Acknowledgments}

This work is partly supported by the National Natural Science Foundation
of China under Grants No. 11375024,  No.11522539, No.11735003,
the Thousand Talents Plan for Young Professionals and
the Fundamental Research Funds for the Central Universities.


\begin{thebibliography}{48}%
\makeatletter
\providecommand \@ifxundefined [1]{%
 \@ifx{#1\undefined}
}%
\providecommand \@ifnum [1]{%
 \ifnum #1\expandafter \@firstoftwo
 \else \expandafter \@secondoftwo
 \fi
}%
\providecommand \@ifx [1]{%
 \ifx #1\expandafter \@firstoftwo
 \else \expandafter \@secondoftwo
 \fi
}%
\providecommand \natexlab [1]{#1}%
\providecommand \enquote  [1]{``#1''}%
\providecommand \bibnamefont  [1]{#1}%
\providecommand \bibfnamefont [1]{#1}%
\providecommand \citenamefont [1]{#1}%
\providecommand \href@noop [0]{\@secondoftwo}%
\providecommand \href [0]{\begingroup \@sanitize@url \@href}%
\providecommand \@href[1]{\@@startlink{#1}\@@href}%
\providecommand \@@href[1]{\endgroup#1\@@endlink}%
\providecommand \@sanitize@url [0]{\catcode `\\12\catcode `\$12\catcode
  `\&12\catcode `\#12\catcode `\^12\catcode `\_12\catcode `\%12\relax}%
\providecommand \@@startlink[1]{}%
\providecommand \@@endlink[0]{}%
\providecommand \url  [0]{\begingroup\@sanitize@url \@url }%
\providecommand \@url [1]{\endgroup\@href {#1}{\urlprefix }}%
\providecommand \urlprefix  [0]{URL }%
\providecommand \Eprint [0]{\href }%
\providecommand \doibase [0]{http://dx.doi.org/}%
\providecommand \selectlanguage [0]{\@gobble}%
\providecommand \bibinfo  [0]{\@secondoftwo}%
\providecommand \bibfield  [0]{\@secondoftwo}%
\providecommand \translation [1]{[#1]}%
\providecommand \BibitemOpen [0]{}%
\providecommand \bibitemStop [0]{}%
\providecommand \bibitemNoStop [0]{.\EOS\space}%
\providecommand \EOS [0]{\spacefactor3000\relax}%
\providecommand \BibitemShut  [1]{\csname bibitem#1\endcsname}%
\let\auto@bib@innerbib\@empty
\bibitem [{\citenamefont {Efimov}(1970)}]{Efimov:1970zz}%
  \BibitemOpen
  \bibfield  {author} {\bibinfo {author} {\bibfnamefont {V.}~\bibnamefont
  {Efimov}},\ }\href {\doibase 10.1016/0370-2693(70)90349-7} {\bibfield
  {journal} {\bibinfo  {journal} {Phys. Lett.}\ }\textbf {\bibinfo {volume}
  {33B}},\ \bibinfo {pages} {563} (\bibinfo {year} {1970})}\BibitemShut
  {NoStop}%
\bibitem [{\citenamefont {Kraemer}\ \emph {et~al.}(2006)\citenamefont
  {Kraemer}, \citenamefont {Mark}, \citenamefont {Waldburger}, \citenamefont
  {Danzl}, \citenamefont {Chin}, \citenamefont {Engeser}, \citenamefont
  {Lange}, \citenamefont {Pilch}, \citenamefont {Jaakkola}, \citenamefont
  {Nägerl},\ and\ \citenamefont {Grimm}}]{Kraemer:2006}%
  \BibitemOpen
  \bibfield  {author} {\bibinfo {author} {\bibfnamefont {T.}~\bibnamefont
  {Kraemer}}, \bibinfo {author} {\bibfnamefont {M.}~\bibnamefont {Mark}},
  \bibinfo {author} {\bibfnamefont {P.}~\bibnamefont {Waldburger}}, \bibinfo
  {author} {\bibfnamefont {J.~G.}\ \bibnamefont {Danzl}}, \bibinfo {author}
  {\bibfnamefont {C.}~\bibnamefont {Chin}}, \bibinfo {author} {\bibfnamefont
  {B.}~\bibnamefont {Engeser}}, \bibinfo {author} {\bibfnamefont {A.~D.}\
  \bibnamefont {Lange}}, \bibinfo {author} {\bibfnamefont {K.}~\bibnamefont
  {Pilch}}, \bibinfo {author} {\bibfnamefont {A.}~\bibnamefont {Jaakkola}},
  \bibinfo {author} {\bibfnamefont {H.-C.}\ \bibnamefont {Nägerl}}, \ and\
  \bibinfo {author} {\bibfnamefont {R.}~\bibnamefont {Grimm}},\ }\href
  {\doibase 10.1038/nature04626} {\bibfield  {journal} {\bibinfo  {journal}
  {Nature}\ }\textbf {\bibinfo {volume} {440}},\ \bibinfo {pages} {315}
  (\bibinfo {year} {2006})}\BibitemShut {NoStop}%
\bibitem [{\citenamefont {Braaten}\ and\ \citenamefont
  {Hammer}(2006)}]{Braaten:2004rn}%
  \BibitemOpen
  \bibfield  {author} {\bibinfo {author} {\bibfnamefont {E.}~\bibnamefont
  {Braaten}}\ and\ \bibinfo {author} {\bibfnamefont {H.~W.}\ \bibnamefont
  {Hammer}},\ }\href {\doibase 10.1016/j.physrep.2006.03.001} {\bibfield
  {journal} {\bibinfo  {journal} {Phys. Rept.}\ }\textbf {\bibinfo {volume}
  {428}},\ \bibinfo {pages} {259} (\bibinfo {year} {2006})},\ \Eprint
  {http://arxiv.org/abs/cond-mat/0410417} {arXiv:cond-mat/0410417 [cond-mat]}
  \BibitemShut {NoStop}%
\bibitem [{\citenamefont {Naidon}\ and\ \citenamefont
  {Endo}(2017)}]{Naidon:2016dpf}%
  \BibitemOpen
  \bibfield  {author} {\bibinfo {author} {\bibfnamefont {P.}~\bibnamefont
  {Naidon}}\ and\ \bibinfo {author} {\bibfnamefont {S.}~\bibnamefont {Endo}},\
  }\href {\doibase 10.1088/1361-6633/aa50e8} {\bibfield  {journal} {\bibinfo
  {journal} {Rept. Prog. Phys.}\ }\textbf {\bibinfo {volume} {80}},\ \bibinfo
  {pages} {056001} (\bibinfo {year} {2017})},\ \Eprint
  {http://arxiv.org/abs/1610.09805} {arXiv:1610.09805 [quant-ph]} \BibitemShut
  {NoStop}%
\bibitem [{\citenamefont {Geng}\ \emph {et~al.}(2018)\citenamefont {Geng},
  \citenamefont {Lu},\ and\ \citenamefont {Valderrama}}]{Geng:2017hxc}%
  \BibitemOpen
  \bibfield  {author} {\bibinfo {author} {\bibfnamefont {L.}~\bibnamefont
  {Geng}}, \bibinfo {author} {\bibfnamefont {J.}~\bibnamefont {Lu}}, \ and\
  \bibinfo {author} {\bibfnamefont {M.~P.}\ \bibnamefont {Valderrama}},\ }\href
  {\doibase 10.1103/PhysRevD.97.094036} {\bibfield  {journal} {\bibinfo
  {journal} {Phys. Rev.}\ }\textbf {\bibinfo {volume} {D97}},\ \bibinfo {pages}
  {094036} (\bibinfo {year} {2018})},\ \Eprint
  {http://arxiv.org/abs/1704.06123} {arXiv:1704.06123 [hep-ph]} \BibitemShut
  {NoStop}%
\bibitem [{\citenamefont {Pires}\ \emph {et~al.}(2014)\citenamefont {Pires},
  \citenamefont {Ulmanis}, \citenamefont {Häfner}, \citenamefont {Repp},
  \citenamefont {Arias}, \citenamefont {Kuhnle},\ and\ \citenamefont
  {Weidemüller}}]{Pires:2014zza}%
  \BibitemOpen
  \bibfield  {author} {\bibinfo {author} {\bibfnamefont {R.}~\bibnamefont
  {Pires}}, \bibinfo {author} {\bibfnamefont {J.}~\bibnamefont {Ulmanis}},
  \bibinfo {author} {\bibfnamefont {S.}~\bibnamefont {Häfner}}, \bibinfo
  {author} {\bibfnamefont {M.}~\bibnamefont {Repp}}, \bibinfo {author}
  {\bibfnamefont {A.}~\bibnamefont {Arias}}, \bibinfo {author} {\bibfnamefont
  {E.~D.}\ \bibnamefont {Kuhnle}}, \ and\ \bibinfo {author} {\bibfnamefont
  {M.}~\bibnamefont {Weidemüller}},\ }\href {\doibase
  10.1103/PhysRevLett.112.250404} {\bibfield  {journal} {\bibinfo  {journal}
  {Phys. Rev. Lett.}\ }\textbf {\bibinfo {volume} {112}},\ \bibinfo {pages}
  {250404} (\bibinfo {year} {2014})},\ \Eprint {http://arxiv.org/abs/1403.7246}
  {arXiv:1403.7246 [cond-mat.quant-gas]} \BibitemShut {NoStop}%
\bibitem [{\citenamefont {Helfrich}\ and\ \citenamefont
  {Hammer}(2011)}]{Helfrich:2011ut}%
  \BibitemOpen
  \bibfield  {author} {\bibinfo {author} {\bibfnamefont {K.}~\bibnamefont
  {Helfrich}}\ and\ \bibinfo {author} {\bibfnamefont {H.~W.}\ \bibnamefont
  {Hammer}},\ }\href {\doibase 10.1088/0953-4075/44/21/215301} {\bibfield
  {journal} {\bibinfo  {journal} {J. Phys.}\ }\textbf {\bibinfo {volume}
  {B44}},\ \bibinfo {pages} {215301} (\bibinfo {year} {2011})},\ \Eprint
  {http://arxiv.org/abs/1107.0869} {arXiv:1107.0869 [cond-mat.quant-gas]}
  \BibitemShut {NoStop}%
\bibitem [{\citenamefont {Kartavtsev}\ and\ \citenamefont
  {Malykh}(2007)}]{Kartavtsev:2006aa}%
  \BibitemOpen
  \bibfield  {author} {\bibinfo {author} {\bibfnamefont {O.~I.}\ \bibnamefont
  {Kartavtsev}}\ and\ \bibinfo {author} {\bibfnamefont {A.~V.}\ \bibnamefont
  {Malykh}},\ }\href {\doibase 10.1088/0953-4075/40/7/011} {\bibfield
  {journal} {\bibinfo  {journal} {Journal of Physics}\ }\textbf {\bibinfo
  {volume} {B40}},\ \bibinfo {pages} {1429} (\bibinfo {year}
  {2007})}\BibitemShut {NoStop}%
\bibitem [{\citenamefont {Choi}\ \emph {et~al.}(2003)\citenamefont {Choi} \emph
  {et~al.}}]{Choi:2003ue}%
  \BibitemOpen
  \bibfield  {author} {\bibinfo {author} {\bibfnamefont {S.~K.}\ \bibnamefont
  {Choi}} \emph {et~al.} (\bibinfo {collaboration} {Belle}),\ }\href {\doibase
  10.1103/PhysRevLett.91.262001} {\bibfield  {journal} {\bibinfo  {journal}
  {Phys. Rev. Lett.}\ }\textbf {\bibinfo {volume} {91}},\ \bibinfo {pages}
  {262001} (\bibinfo {year} {2003})},\ \Eprint
  {http://arxiv.org/abs/hep-ex/0309032} {arXiv:hep-ex/0309032} \BibitemShut
  {NoStop}%
\bibitem [{\citenamefont {Tornqvist}(2003)}]{Tornqvist:2003na}%
  \BibitemOpen
  \bibfield  {author} {\bibinfo {author} {\bibfnamefont {N.~A.}\ \bibnamefont
  {Tornqvist}},\ }\href@noop {} {\  (\bibinfo {year} {2003})},\ \Eprint
  {http://arxiv.org/abs/hep-ph/0308277} {arXiv:hep-ph/0308277 [hep-ph]}
  \BibitemShut {NoStop}%
\bibitem [{\citenamefont {Voloshin}(2004)}]{Voloshin:2003nt}%
  \BibitemOpen
  \bibfield  {author} {\bibinfo {author} {\bibfnamefont {M.}~\bibnamefont
  {Voloshin}},\ }\href {\doibase 10.1016/j.physletb.2003.11.014} {\bibfield
  {journal} {\bibinfo  {journal} {Phys.Lett.}\ }\textbf {\bibinfo {volume}
  {B579}},\ \bibinfo {pages} {316} (\bibinfo {year} {2004})},\ \Eprint
  {http://arxiv.org/abs/hep-ph/0309307} {arXiv:hep-ph/0309307 [hep-ph]}
  \BibitemShut {NoStop}%
\bibitem [{\citenamefont {Braaten}\ and\ \citenamefont
  {Kusunoki}(2004)}]{Braaten:2003he}%
  \BibitemOpen
  \bibfield  {author} {\bibinfo {author} {\bibfnamefont {E.}~\bibnamefont
  {Braaten}}\ and\ \bibinfo {author} {\bibfnamefont {M.}~\bibnamefont
  {Kusunoki}},\ }\href {\doibase 10.1103/PhysRevD.69.074005} {\bibfield
  {journal} {\bibinfo  {journal} {Phys.Rev.}\ }\textbf {\bibinfo {volume}
  {D69}},\ \bibinfo {pages} {074005} (\bibinfo {year} {2004})},\ \Eprint
  {http://arxiv.org/abs/hep-ph/0311147} {arXiv:hep-ph/0311147 [hep-ph]}
  \BibitemShut {NoStop}%
\bibitem [{\citenamefont {Kolomeitsev}\ and\ \citenamefont
  {Lutz}(2004)}]{Kolomeitsev:2003ac}%
  \BibitemOpen
  \bibfield  {author} {\bibinfo {author} {\bibfnamefont {E.~E.}\ \bibnamefont
  {Kolomeitsev}}\ and\ \bibinfo {author} {\bibfnamefont {M.~F.~M.}\
  \bibnamefont {Lutz}},\ }\href {\doibase 10.1016/j.physletb.2003.10.118}
  {\bibfield  {journal} {\bibinfo  {journal} {Phys. Lett.}\ }\textbf {\bibinfo
  {volume} {B582}},\ \bibinfo {pages} {39} (\bibinfo {year} {2004})},\ \Eprint
  {http://arxiv.org/abs/hep-ph/0307133} {arXiv:hep-ph/0307133 [hep-ph]}
  \BibitemShut {NoStop}%
\bibitem [{\citenamefont {Guo}\ \emph {et~al.}(2006)\citenamefont {Guo},
  \citenamefont {Shen}, \citenamefont {Chiang}, \citenamefont {Ping},\ and\
  \citenamefont {Zou}}]{Guo:2006fu}%
  \BibitemOpen
  \bibfield  {author} {\bibinfo {author} {\bibfnamefont {F.-K.}\ \bibnamefont
  {Guo}}, \bibinfo {author} {\bibfnamefont {P.-N.}\ \bibnamefont {Shen}},
  \bibinfo {author} {\bibfnamefont {H.-C.}\ \bibnamefont {Chiang}}, \bibinfo
  {author} {\bibfnamefont {R.-G.}\ \bibnamefont {Ping}}, \ and\ \bibinfo
  {author} {\bibfnamefont {B.-S.}\ \bibnamefont {Zou}},\ }\href {\doibase
  10.1016/j.physletb.2006.08.064} {\bibfield  {journal} {\bibinfo  {journal}
  {Phys. Lett.}\ }\textbf {\bibinfo {volume} {B641}},\ \bibinfo {pages} {278}
  (\bibinfo {year} {2006})},\ \Eprint {http://arxiv.org/abs/hep-ph/0603072}
  {arXiv:hep-ph/0603072 [hep-ph]} \BibitemShut {NoStop}%
\bibitem [{\citenamefont {Guo}\ \emph {et~al.}(2007)\citenamefont {Guo},
  \citenamefont {Shen},\ and\ \citenamefont {Chiang}}]{Guo:2006rp}%
  \BibitemOpen
  \bibfield  {author} {\bibinfo {author} {\bibfnamefont {F.-K.}\ \bibnamefont
  {Guo}}, \bibinfo {author} {\bibfnamefont {P.-N.}\ \bibnamefont {Shen}}, \
  and\ \bibinfo {author} {\bibfnamefont {H.-C.}\ \bibnamefont {Chiang}},\
  }\href {\doibase 10.1016/j.physletb.2007.01.050} {\bibfield  {journal}
  {\bibinfo  {journal} {Phys. Lett.}\ }\textbf {\bibinfo {volume} {B647}},\
  \bibinfo {pages} {133} (\bibinfo {year} {2007})},\ \Eprint
  {http://arxiv.org/abs/hep-ph/0610008} {arXiv:hep-ph/0610008 [hep-ph]}
  \BibitemShut {NoStop}%
\bibitem [{\citenamefont {Guo}\ \emph {et~al.}(2009)\citenamefont {Guo},
  \citenamefont {Hanhart},\ and\ \citenamefont {Meissner}}]{Guo:2009ct}%
  \BibitemOpen
  \bibfield  {author} {\bibinfo {author} {\bibfnamefont {F.-K.}\ \bibnamefont
  {Guo}}, \bibinfo {author} {\bibfnamefont {C.}~\bibnamefont {Hanhart}}, \ and\
  \bibinfo {author} {\bibfnamefont {U.-G.}\ \bibnamefont {Meissner}},\ }\href
  {\doibase 10.1140/epja/i2009-10762-1} {\bibfield  {journal} {\bibinfo
  {journal} {Eur. Phys. J.}\ }\textbf {\bibinfo {volume} {A40}},\ \bibinfo
  {pages} {171} (\bibinfo {year} {2009})},\ \Eprint
  {http://arxiv.org/abs/0901.1597} {arXiv:0901.1597 [hep-ph]} \BibitemShut
  {NoStop}%
\bibitem [{\citenamefont {Altenbuchinger}\ \emph {et~al.}(2014)\citenamefont
  {Altenbuchinger}, \citenamefont {Geng},\ and\ \citenamefont
  {Weise}}]{Altenbuchinger:2013vwa}%
  \BibitemOpen
  \bibfield  {author} {\bibinfo {author} {\bibfnamefont {M.}~\bibnamefont
  {Altenbuchinger}}, \bibinfo {author} {\bibfnamefont {L.~S.}\ \bibnamefont
  {Geng}}, \ and\ \bibinfo {author} {\bibfnamefont {W.}~\bibnamefont {Weise}},\
  }\href {\doibase 10.1103/PhysRevD.89.014026} {\bibfield  {journal} {\bibinfo
  {journal} {Phys. Rev.}\ }\textbf {\bibinfo {volume} {D89}},\ \bibinfo {pages}
  {014026} (\bibinfo {year} {2014})},\ \Eprint {http://arxiv.org/abs/1309.4743}
  {arXiv:1309.4743 [hep-ph]} \BibitemShut {NoStop}%
\bibitem [{\citenamefont {Martínez~Torres}\ \emph {et~al.}(2015)\citenamefont
  {Martínez~Torres}, \citenamefont {Oset}, \citenamefont {Prelovsek},\ and\
  \citenamefont {Ramos}}]{Torres:2014vna}%
  \BibitemOpen
  \bibfield  {author} {\bibinfo {author} {\bibfnamefont {A.}~\bibnamefont
  {Martínez~Torres}}, \bibinfo {author} {\bibfnamefont {E.}~\bibnamefont
  {Oset}}, \bibinfo {author} {\bibfnamefont {S.}~\bibnamefont {Prelovsek}}, \
  and\ \bibinfo {author} {\bibfnamefont {A.}~\bibnamefont {Ramos}},\ }\href
  {\doibase 10.1007/JHEP05(2015)153} {\bibfield  {journal} {\bibinfo  {journal}
  {JHEP}\ }\textbf {\bibinfo {volume} {05}},\ \bibinfo {pages} {153} (\bibinfo
  {year} {2015})},\ \Eprint {http://arxiv.org/abs/1412.1706} {arXiv:1412.1706
  [hep-lat]} \BibitemShut {NoStop}%
\bibitem [{\citenamefont {Bali}\ \emph {et~al.}(2017)\citenamefont {Bali},
  \citenamefont {Collins}, \citenamefont {Cox},\ and\ \citenamefont
  {Schäfer}}]{Bali:2017pdv}%
  \BibitemOpen
  \bibfield  {author} {\bibinfo {author} {\bibfnamefont {G.~S.}\ \bibnamefont
  {Bali}}, \bibinfo {author} {\bibfnamefont {S.}~\bibnamefont {Collins}},
  \bibinfo {author} {\bibfnamefont {A.}~\bibnamefont {Cox}}, \ and\ \bibinfo
  {author} {\bibfnamefont {A.}~\bibnamefont {Schäfer}},\ }\href {\doibase
  10.1103/PhysRevD.96.074501} {\bibfield  {journal} {\bibinfo  {journal} {Phys.
  Rev.}\ }\textbf {\bibinfo {volume} {D96}},\ \bibinfo {pages} {074501}
  (\bibinfo {year} {2017})},\ \Eprint {http://arxiv.org/abs/1706.01247}
  {arXiv:1706.01247 [hep-lat]} \BibitemShut {NoStop}%
\bibitem [{\citenamefont {Savage}\ and\ \citenamefont
  {Wise}(1990)}]{Savage:1990di}%
  \BibitemOpen
  \bibfield  {author} {\bibinfo {author} {\bibfnamefont {M.~J.}\ \bibnamefont
  {Savage}}\ and\ \bibinfo {author} {\bibfnamefont {M.~B.}\ \bibnamefont
  {Wise}},\ }\href {\doibase 10.1016/0370-2693(90)90035-5} {\bibfield
  {journal} {\bibinfo  {journal} {Phys. Lett.}\ }\textbf {\bibinfo {volume}
  {B248}},\ \bibinfo {pages} {177} (\bibinfo {year} {1990})}\BibitemShut
  {NoStop}%
\bibitem [{\citenamefont {Hu}\ and\ \citenamefont {Mehen}(2006)}]{Hu:2005gf}%
  \BibitemOpen
  \bibfield  {author} {\bibinfo {author} {\bibfnamefont {J.}~\bibnamefont
  {Hu}}\ and\ \bibinfo {author} {\bibfnamefont {T.}~\bibnamefont {Mehen}},\
  }\href {\doibase 10.1103/PhysRevD.73.054003} {\bibfield  {journal} {\bibinfo
  {journal} {Phys. Rev.}\ }\textbf {\bibinfo {volume} {D73}},\ \bibinfo {pages}
  {054003} (\bibinfo {year} {2006})},\ \Eprint
  {http://arxiv.org/abs/hep-ph/0511321} {arXiv:hep-ph/0511321 [hep-ph]}
  \BibitemShut {NoStop}%
\bibitem [{\citenamefont {Guo}\ \emph {et~al.}(2013)\citenamefont {Guo},
  \citenamefont {Hidalgo-Duque}, \citenamefont {Nieves},\ and\ \citenamefont
  {Valderrama}}]{Guo:2013xga}%
  \BibitemOpen
  \bibfield  {author} {\bibinfo {author} {\bibfnamefont {F.-K.}\ \bibnamefont
  {Guo}}, \bibinfo {author} {\bibfnamefont {C.}~\bibnamefont {Hidalgo-Duque}},
  \bibinfo {author} {\bibfnamefont {J.}~\bibnamefont {Nieves}}, \ and\ \bibinfo
  {author} {\bibfnamefont {M.~P.}\ \bibnamefont {Valderrama}},\ }\href
  {\doibase 10.1103/PhysRevD.88.054014} {\bibfield  {journal} {\bibinfo
  {journal} {Phys. Rev.}\ }\textbf {\bibinfo {volume} {D88}},\ \bibinfo {pages}
  {054014} (\bibinfo {year} {2013})},\ \Eprint {http://arxiv.org/abs/1305.4052}
  {arXiv:1305.4052 [hep-ph]} \BibitemShut {NoStop}%
\bibitem [{\citenamefont {Guo}\ and\ \citenamefont
  {Meissner}(2011)}]{Guo:2011dd}%
  \BibitemOpen
  \bibfield  {author} {\bibinfo {author} {\bibfnamefont {F.-K.}\ \bibnamefont
  {Guo}}\ and\ \bibinfo {author} {\bibfnamefont {U.-G.}\ \bibnamefont
  {Meissner}},\ }\href {\doibase 10.1103/PhysRevD.84.014013} {\bibfield
  {journal} {\bibinfo  {journal} {Phys. Rev.}\ }\textbf {\bibinfo {volume}
  {D84}},\ \bibinfo {pages} {014013} (\bibinfo {year} {2011})},\ \Eprint
  {http://arxiv.org/abs/1102.3536} {arXiv:1102.3536 [hep-ph]} \BibitemShut
  {NoStop}%
\bibitem [{\citenamefont {Lewis}\ and\ \citenamefont
  {Woloshyn}(2009)}]{Lewis:2008fu}%
  \BibitemOpen
  \bibfield  {author} {\bibinfo {author} {\bibfnamefont {R.}~\bibnamefont
  {Lewis}}\ and\ \bibinfo {author} {\bibfnamefont {R.~M.}\ \bibnamefont
  {Woloshyn}},\ }\href {\doibase 10.1103/PhysRevD.79.014502} {\bibfield
  {journal} {\bibinfo  {journal} {Phys. Rev.}\ }\textbf {\bibinfo {volume}
  {D79}},\ \bibinfo {pages} {014502} (\bibinfo {year} {2009})},\ \Eprint
  {http://arxiv.org/abs/0806.4783} {arXiv:0806.4783 [hep-lat]} \BibitemShut
  {NoStop}%
\bibitem [{\citenamefont {Sanchez~Sanchez}\ \emph {et~al.}(2017)\citenamefont
  {Sanchez~Sanchez}, \citenamefont {Geng}, \citenamefont {Lu}, \citenamefont
  {Hyodo},\ and\ \citenamefont {Valderrama}}]{SanchezSanchez:2017xtl}%
  \BibitemOpen
  \bibfield  {author} {\bibinfo {author} {\bibfnamefont {M.}~\bibnamefont
  {Sanchez~Sanchez}}, \bibinfo {author} {\bibfnamefont {L.-S.}\ \bibnamefont
  {Geng}}, \bibinfo {author} {\bibfnamefont {J.-X.}\ \bibnamefont {Lu}},
  \bibinfo {author} {\bibfnamefont {T.}~\bibnamefont {Hyodo}}, \ and\ \bibinfo
  {author} {\bibfnamefont {M.~P.}\ \bibnamefont {Valderrama}},\ }\href@noop {}
  {\  (\bibinfo {year} {2017})},\ \Eprint {http://arxiv.org/abs/1707.03802}
  {arXiv:1707.03802 [hep-ph]} \BibitemShut {NoStop}%
\bibitem [{\citenamefont {Bayar}\ \emph {et~al.}(2015)\citenamefont {Bayar},
  \citenamefont {Ren},\ and\ \citenamefont {Oset}}]{Bayar:2015oea}%
  \BibitemOpen
  \bibfield  {author} {\bibinfo {author} {\bibfnamefont {M.}~\bibnamefont
  {Bayar}}, \bibinfo {author} {\bibfnamefont {X.-L.}\ \bibnamefont {Ren}}, \
  and\ \bibinfo {author} {\bibfnamefont {E.}~\bibnamefont {Oset}},\ }\href
  {\doibase 10.1140/epja/i2015-15061-8} {\bibfield  {journal} {\bibinfo
  {journal} {Eur. Phys. J.}\ }\textbf {\bibinfo {volume} {A51}},\ \bibinfo
  {pages} {61} (\bibinfo {year} {2015})},\ \Eprint
  {http://arxiv.org/abs/1501.02962} {arXiv:1501.02962 [hep-ph]} \BibitemShut
  {NoStop}%
\bibitem [{\citenamefont {Bayar}\ \emph {et~al.}(2016)\citenamefont {Bayar},
  \citenamefont {Fernandez-Soler}, \citenamefont {Sun},\ and\ \citenamefont
  {Oset}}]{Bayar:2015zba}%
  \BibitemOpen
  \bibfield  {author} {\bibinfo {author} {\bibfnamefont {M.}~\bibnamefont
  {Bayar}}, \bibinfo {author} {\bibfnamefont {P.}~\bibnamefont
  {Fernandez-Soler}}, \bibinfo {author} {\bibfnamefont {Z.-F.}\ \bibnamefont
  {Sun}}, \ and\ \bibinfo {author} {\bibfnamefont {E.}~\bibnamefont {Oset}},\
  }\href {\doibase 10.1140/epja/i2016-16106-2} {\bibfield  {journal} {\bibinfo
  {journal} {Eur. Phys. J.}\ }\textbf {\bibinfo {volume} {A52}},\ \bibinfo
  {pages} {106} (\bibinfo {year} {2016})},\ \Eprint
  {http://arxiv.org/abs/1510.06570} {arXiv:1510.06570 [hep-ph]} \BibitemShut
  {NoStop}%
\bibitem [{\citenamefont {Ren}\ \emph {et~al.}(2018)\citenamefont {Ren},
  \citenamefont {Malabarba}, \citenamefont {Geng}, \citenamefont
  {Khemchandani},\ and\ \citenamefont {Martínez~Torres}}]{Ren:2018pcd}%
  \BibitemOpen
  \bibfield  {author} {\bibinfo {author} {\bibfnamefont {X.-L.}\ \bibnamefont
  {Ren}}, \bibinfo {author} {\bibfnamefont {B.~B.}\ \bibnamefont {Malabarba}},
  \bibinfo {author} {\bibfnamefont {L.-S.}\ \bibnamefont {Geng}}, \bibinfo
  {author} {\bibfnamefont {K.~P.}\ \bibnamefont {Khemchandani}}, \ and\
  \bibinfo {author} {\bibfnamefont {A.}~\bibnamefont {Martínez~Torres}},\
  }\href@noop {} {\  (\bibinfo {year} {2018})},\ \Eprint
  {http://arxiv.org/abs/1805.08330} {arXiv:1805.08330 [hep-ph]} \BibitemShut
  {NoStop}%
\bibitem [{\citenamefont {Lu}\ \emph {et~al.}(2015)\citenamefont {Lu},
  \citenamefont {Zhou}, \citenamefont {Chen}, \citenamefont {Xie},\ and\
  \citenamefont {Geng}}]{Lu:2014ina}%
  \BibitemOpen
  \bibfield  {author} {\bibinfo {author} {\bibfnamefont {J.-X.}\ \bibnamefont
  {Lu}}, \bibinfo {author} {\bibfnamefont {Y.}~\bibnamefont {Zhou}}, \bibinfo
  {author} {\bibfnamefont {H.-X.}\ \bibnamefont {Chen}}, \bibinfo {author}
  {\bibfnamefont {J.-J.}\ \bibnamefont {Xie}}, \ and\ \bibinfo {author}
  {\bibfnamefont {L.-S.}\ \bibnamefont {Geng}},\ }\href {\doibase
  10.1103/PhysRevD.92.014036} {\bibfield  {journal} {\bibinfo  {journal} {Phys.
  Rev.}\ }\textbf {\bibinfo {volume} {D92}},\ \bibinfo {pages} {014036}
  (\bibinfo {year} {2015})},\ \Eprint {http://arxiv.org/abs/1409.3133}
  {arXiv:1409.3133 [hep-ph]} \BibitemShut {NoStop}%
\bibitem [{\citenamefont {Aaij}\ \emph {et~al.}(2017)\citenamefont {Aaij} \emph
  {et~al.}}]{Aaij:2017ueg}%
  \BibitemOpen
  \bibfield  {author} {\bibinfo {author} {\bibfnamefont {R.}~\bibnamefont
  {Aaij}} \emph {et~al.} (\bibinfo {collaboration} {LHCb}),\ }\href {\doibase
  10.1103/PhysRevLett.119.112001} {\bibfield  {journal} {\bibinfo  {journal}
  {Phys. Rev. Lett.}\ }\textbf {\bibinfo {volume} {119}},\ \bibinfo {pages}
  {112001} (\bibinfo {year} {2017})},\ \Eprint
  {http://arxiv.org/abs/1707.01621} {arXiv:1707.01621 [hep-ex]} \BibitemShut
  {NoStop}%
\bibitem [{\citenamefont {Karliner}\ and\ \citenamefont
  {Rosner}(2017)}]{Karliner:2017qjm}%
  \BibitemOpen
  \bibfield  {author} {\bibinfo {author} {\bibfnamefont {M.}~\bibnamefont
  {Karliner}}\ and\ \bibinfo {author} {\bibfnamefont {J.~L.}\ \bibnamefont
  {Rosner}},\ }\href {\doibase 10.1103/PhysRevLett.119.202001} {\bibfield
  {journal} {\bibinfo  {journal} {Phys. Rev. Lett.}\ }\textbf {\bibinfo
  {volume} {119}},\ \bibinfo {pages} {202001} (\bibinfo {year} {2017})},\
  \Eprint {http://arxiv.org/abs/1707.07666} {arXiv:1707.07666 [hep-ph]}
  \BibitemShut {NoStop}%
\bibitem [{\citenamefont {Mehen}(2017)}]{Mehen:2017nrh}%
  \BibitemOpen
  \bibfield  {author} {\bibinfo {author} {\bibfnamefont {T.}~\bibnamefont
  {Mehen}},\ }\href {\doibase 10.1103/PhysRevD.96.094028} {\bibfield  {journal}
  {\bibinfo  {journal} {Phys. Rev.}\ }\textbf {\bibinfo {volume} {D96}},\
  \bibinfo {pages} {094028} (\bibinfo {year} {2017})},\ \Eprint
  {http://arxiv.org/abs/1708.05020} {arXiv:1708.05020 [hep-ph]} \BibitemShut
  {NoStop}%
\bibitem [{\citenamefont {Tornqvist}(1994)}]{Tornqvist:1993ng}%
  \BibitemOpen
  \bibfield  {author} {\bibinfo {author} {\bibfnamefont {N.~A.}\ \bibnamefont
  {Tornqvist}},\ }\href {\doibase 10.1007/BF01413192} {\bibfield  {journal}
  {\bibinfo  {journal} {Z.Phys.}\ }\textbf {\bibinfo {volume} {C61}},\ \bibinfo
  {pages} {525} (\bibinfo {year} {1994})},\ \Eprint
  {http://arxiv.org/abs/hep-ph/9310247} {arXiv:hep-ph/9310247 [hep-ph]}
  \BibitemShut {NoStop}%
\bibitem [{\citenamefont {Liu}\ \emph {et~al.}(2018)\citenamefont {Liu},
  \citenamefont {Wu}, \citenamefont {Xie}, \citenamefont {Pavon~Valderrama},\
  and\ \citenamefont {Geng}}]{Liu:2018bkx}%
  \BibitemOpen
  \bibfield  {author} {\bibinfo {author} {\bibfnamefont {M.-Z.}\ \bibnamefont
  {Liu}}, \bibinfo {author} {\bibfnamefont {T.-W.}\ \bibnamefont {Wu}},
  \bibinfo {author} {\bibfnamefont {J.-J.}\ \bibnamefont {Xie}}, \bibinfo
  {author} {\bibfnamefont {M.}~\bibnamefont {Pavon~Valderrama}}, \ and\
  \bibinfo {author} {\bibfnamefont {L.-S.}\ \bibnamefont {Geng}},\ }\href
  {\doibase 10.1103/PhysRevD.98.014014} {\bibfield  {journal} {\bibinfo
  {journal} {Phys. Rev.}\ }\textbf {\bibinfo {volume} {D98}},\ \bibinfo {pages}
  {014014} (\bibinfo {year} {2018})},\ \Eprint
  {http://arxiv.org/abs/1805.08384} {arXiv:1805.08384 [hep-ph]} \BibitemShut
  {NoStop}%
\bibitem [{\citenamefont {Kadyshevsky}(1968)}]{Kadyshevsky:1967rs}%
  \BibitemOpen
  \bibfield  {author} {\bibinfo {author} {\bibfnamefont {V.}~\bibnamefont
  {Kadyshevsky}},\ }\href {\doibase 10.1016/0550-3213(68)90274-5} {\bibfield
  {journal} {\bibinfo  {journal} {Nucl.Phys.}\ }\textbf {\bibinfo {volume}
  {B6}},\ \bibinfo {pages} {125} (\bibinfo {year} {1968})}\BibitemShut
  {NoStop}%
\bibitem [{\citenamefont {Gross}(1969)}]{Gross:1969rv}%
  \BibitemOpen
  \bibfield  {author} {\bibinfo {author} {\bibfnamefont {F.}~\bibnamefont
  {Gross}},\ }\href {\doibase 10.1103/PhysRev.186.1448} {\bibfield  {journal}
  {\bibinfo  {journal} {Phys.Rev.}\ }\textbf {\bibinfo {volume} {186}},\
  \bibinfo {pages} {1448} (\bibinfo {year} {1969})}\BibitemShut {NoStop}%
\bibitem [{\citenamefont {Yaes}(1971)}]{Yaes:1971vw}%
  \BibitemOpen
  \bibfield  {author} {\bibinfo {author} {\bibfnamefont {R.}~\bibnamefont
  {Yaes}},\ }\href {\doibase 10.1103/PhysRevD.3.3086} {\bibfield  {journal}
  {\bibinfo  {journal} {Phys.Rev.}\ }\textbf {\bibinfo {volume} {D3}},\
  \bibinfo {pages} {3086} (\bibinfo {year} {1971})}\BibitemShut {NoStop}%
\bibitem [{\citenamefont {Woloshyn}\ and\ \citenamefont
  {Jackson}(1973)}]{Woloshyn:1974wm}%
  \BibitemOpen
  \bibfield  {author} {\bibinfo {author} {\bibfnamefont {R.}~\bibnamefont
  {Woloshyn}}\ and\ \bibinfo {author} {\bibfnamefont {A.}~\bibnamefont
  {Jackson}},\ }\href {\doibase 10.1016/0550-3213(73)90626-3} {\bibfield
  {journal} {\bibinfo  {journal} {Nucl.Phys.}\ }\textbf {\bibinfo {volume}
  {B64}},\ \bibinfo {pages} {269} (\bibinfo {year} {1973})}\BibitemShut
  {NoStop}%
\bibitem [{\citenamefont {Ramalho}\ \emph {et~al.}(2002)\citenamefont
  {Ramalho}, \citenamefont {Arriaga},\ and\ \citenamefont
  {Pena}}]{Ramalho:2001pd}%
  \BibitemOpen
  \bibfield  {author} {\bibinfo {author} {\bibfnamefont {G.}~\bibnamefont
  {Ramalho}}, \bibinfo {author} {\bibfnamefont {A.}~\bibnamefont {Arriaga}}, \
  and\ \bibinfo {author} {\bibfnamefont {M.}~\bibnamefont {Pena}},\ }\href
  {\doibase 10.1103/PhysRevC.65.034008} {\bibfield  {journal} {\bibinfo
  {journal} {Phys.Rev.}\ }\textbf {\bibinfo {volume} {C65}},\ \bibinfo {pages}
  {034008} (\bibinfo {year} {2002})},\ \Eprint
  {http://arxiv.org/abs/nucl-th/0107065} {arXiv:nucl-th/0107065 [nucl-th]}
  \BibitemShut {NoStop}%
\bibitem [{\citenamefont {Garcilazo}\ and\ \citenamefont
  {Mathelitsch}(1983)}]{Garcilazo:1984rx}%
  \BibitemOpen
  \bibfield  {author} {\bibinfo {author} {\bibfnamefont {H.}~\bibnamefont
  {Garcilazo}}\ and\ \bibinfo {author} {\bibfnamefont {L.}~\bibnamefont
  {Mathelitsch}},\ }\href {\doibase 10.1103/PhysRevC.28.1272} {\bibfield
  {journal} {\bibinfo  {journal} {Phys. Rev.}\ }\textbf {\bibinfo {volume}
  {C28}},\ \bibinfo {pages} {1272} (\bibinfo {year} {1983})}\BibitemShut
  {NoStop}%
\bibitem [{\citenamefont {Mathelitsch}\ and\ \citenamefont
  {Garcilazo}(1985)}]{Mathelitsch:1986ez}%
  \BibitemOpen
  \bibfield  {author} {\bibinfo {author} {\bibfnamefont {L.}~\bibnamefont
  {Mathelitsch}}\ and\ \bibinfo {author} {\bibfnamefont {H.}~\bibnamefont
  {Garcilazo}},\ }\href {\doibase 10.1103/PhysRevC.32.1635} {\bibfield
  {journal} {\bibinfo  {journal} {Phys. Rev.}\ }\textbf {\bibinfo {volume}
  {C32}},\ \bibinfo {pages} {1635} (\bibinfo {year} {1985})}\BibitemShut
  {NoStop}%
\bibitem [{\citenamefont {Hyodo}\ and\ \citenamefont
  {Weise}(2008)}]{Hyodo:2007jq}%
  \BibitemOpen
  \bibfield  {author} {\bibinfo {author} {\bibfnamefont {T.}~\bibnamefont
  {Hyodo}}\ and\ \bibinfo {author} {\bibfnamefont {W.}~\bibnamefont {Weise}},\
  }\href {\doibase 10.1103/PhysRevC.77.035204} {\bibfield  {journal} {\bibinfo
  {journal} {Phys. Rev.}\ }\textbf {\bibinfo {volume} {C77}},\ \bibinfo {pages}
  {035204} (\bibinfo {year} {2008})},\ \Eprint {http://arxiv.org/abs/0712.1613}
  {arXiv:0712.1613 [nucl-th]} \BibitemShut {NoStop}%
\bibitem [{\citenamefont {Patrignani}\ \emph {et~al.}(2016)\citenamefont
  {Patrignani} \emph {et~al.}}]{Patrignani:2016xqp}%
  \BibitemOpen
  \bibfield  {author} {\bibinfo {author} {\bibfnamefont {C.}~\bibnamefont
  {Patrignani}} \emph {et~al.} (\bibinfo {collaboration} {Particle Data
  Group}),\ }\href {\doibase 10.1088/1674-1137/40/10/100001} {\bibfield
  {journal} {\bibinfo  {journal} {Chin. Phys.}\ }\textbf {\bibinfo {volume}
  {C40}},\ \bibinfo {pages} {100001} (\bibinfo {year} {2016})}\BibitemShut
  {NoStop}%
\bibitem [{\citenamefont {Jido}\ \emph {et~al.}(2003)\citenamefont {Jido},
  \citenamefont {Oller}, \citenamefont {Oset}, \citenamefont {Ramos},\ and\
  \citenamefont {Meissner}}]{Jido:2003cb}%
  \BibitemOpen
  \bibfield  {author} {\bibinfo {author} {\bibfnamefont {D.}~\bibnamefont
  {Jido}}, \bibinfo {author} {\bibfnamefont {J.~A.}\ \bibnamefont {Oller}},
  \bibinfo {author} {\bibfnamefont {E.}~\bibnamefont {Oset}}, \bibinfo {author}
  {\bibfnamefont {A.}~\bibnamefont {Ramos}}, \ and\ \bibinfo {author}
  {\bibfnamefont {U.~G.}\ \bibnamefont {Meissner}},\ }\href {\doibase
  10.1016/S0375-9474(03)01598-7} {\bibfield  {journal} {\bibinfo  {journal}
  {Nucl. Phys.}\ }\textbf {\bibinfo {volume} {A725}},\ \bibinfo {pages} {181}
  (\bibinfo {year} {2003})},\ \Eprint {http://arxiv.org/abs/nucl-th/0303062}
  {arXiv:nucl-th/0303062 [nucl-th]} \BibitemShut {NoStop}%
\bibitem [{\citenamefont {Magas}\ \emph {et~al.}(2005)\citenamefont {Magas},
  \citenamefont {Oset},\ and\ \citenamefont {Ramos}}]{Magas:2005vu}%
  \BibitemOpen
  \bibfield  {author} {\bibinfo {author} {\bibfnamefont {V.~K.}\ \bibnamefont
  {Magas}}, \bibinfo {author} {\bibfnamefont {E.}~\bibnamefont {Oset}}, \ and\
  \bibinfo {author} {\bibfnamefont {A.}~\bibnamefont {Ramos}},\ }\href
  {\doibase 10.1103/PhysRevLett.95.052301} {\bibfield  {journal} {\bibinfo
  {journal} {Phys. Rev. Lett.}\ }\textbf {\bibinfo {volume} {95}},\ \bibinfo
  {pages} {052301} (\bibinfo {year} {2005})},\ \Eprint
  {http://arxiv.org/abs/hep-ph/0503043} {arXiv:hep-ph/0503043 [hep-ph]}
  \BibitemShut {NoStop}%
\bibitem [{\citenamefont {Hyodo}\ \emph {et~al.}(2006)\citenamefont {Hyodo},
  \citenamefont {Jido},\ and\ \citenamefont {Hosaka}}]{Hyodo:2006yk}%
  \BibitemOpen
  \bibfield  {author} {\bibinfo {author} {\bibfnamefont {T.}~\bibnamefont
  {Hyodo}}, \bibinfo {author} {\bibfnamefont {D.}~\bibnamefont {Jido}}, \ and\
  \bibinfo {author} {\bibfnamefont {A.}~\bibnamefont {Hosaka}},\ }\href
  {\doibase 10.1103/PhysRevLett.97.192002} {\bibfield  {journal} {\bibinfo
  {journal} {Phys. Rev. Lett.}\ }\textbf {\bibinfo {volume} {97}},\ \bibinfo
  {pages} {192002} (\bibinfo {year} {2006})},\ \Eprint
  {http://arxiv.org/abs/hep-ph/0609014} {arXiv:hep-ph/0609014 [hep-ph]}
  \BibitemShut {NoStop}%
\bibitem [{\citenamefont {Hyodo}\ \emph {et~al.}(2007)\citenamefont {Hyodo},
  \citenamefont {Jido},\ and\ \citenamefont {Hosaka}}]{Hyodo:2006kg}%
  \BibitemOpen
  \bibfield  {author} {\bibinfo {author} {\bibfnamefont {T.}~\bibnamefont
  {Hyodo}}, \bibinfo {author} {\bibfnamefont {D.}~\bibnamefont {Jido}}, \ and\
  \bibinfo {author} {\bibfnamefont {A.}~\bibnamefont {Hosaka}},\ }\href
  {\doibase 10.1103/PhysRevD.75.034002} {\bibfield  {journal} {\bibinfo
  {journal} {Phys. Rev.}\ }\textbf {\bibinfo {volume} {D75}},\ \bibinfo {pages}
  {034002} (\bibinfo {year} {2007})},\ \Eprint
  {http://arxiv.org/abs/hep-ph/0611004} {arXiv:hep-ph/0611004 [hep-ph]}
  \BibitemShut {NoStop}%
\bibitem [{\citenamefont {Karliner}\ and\ \citenamefont
  {Nussinov}(2013)}]{Karliner:2013dqa}%
  \BibitemOpen
  \bibfield  {author} {\bibinfo {author} {\bibfnamefont {M.}~\bibnamefont
  {Karliner}}\ and\ \bibinfo {author} {\bibfnamefont {S.}~\bibnamefont
  {Nussinov}},\ }\href {\doibase 10.1007/JHEP07(2013)153} {\bibfield  {journal}
  {\bibinfo  {journal} {JHEP}\ }\textbf {\bibinfo {volume} {07}},\ \bibinfo
  {pages} {153} (\bibinfo {year} {2013})},\ \Eprint
  {http://arxiv.org/abs/1304.0345} {arXiv:1304.0345 [hep-ph]} \BibitemShut
  {NoStop}%
\end{thebibliography}

%

\end{document}